\documentclass[runningheads]{llncs}
\usepackage{graphicx}
\usepackage{booktabs}
\usepackage{threeparttable}
\begin{document}
\title{Large Batch and Patch Size Training \\for Medical Image Segmentation\thanks{This work was partly achieved through the use of SQUID at the Cybermedia Center, Osaka University. \\This study was supported by JSPS KAKENHI (grant numbers: JP21H03840).}}
%
%

\author{Junya Sato\and
Shoji Kido}
\authorrunning{J. Sato and S. Kido.}
\titlerunning{Large batch and patch size training for medical image segmentation}
%
\institute{Osaka University Graduate School of Medicine, 2-2 Yamadaoka, Suita-city, Osaka 565-0871, Japan \\
\email{\{j-sato,kido\}@radiol.med.osaka-u.ac.jp}}
\maketitle              
\begin{abstract}
Multi-organ segmentation enables organ evaluation, accounts the relationship between multiple organs, and facilitates accurate diagnosis and treatment decisions. However, only few models can perform segmentation accurately because of the lack of datasets and computational resources. On AMOS2022 challenge, which is a large-scale, clinical, and diverse abdominal multiorgan segmentation benchmark, we trained a 3D-UNet model with large batch and patch sizes using multi-GPU distributed training. Segmentation performance tended to increase for models with large batch and patch sizes compared with the baseline settings. The accuracy was further improved by using ensemble models that were trained with different settings. These results provide a reference for parameter selection in organ segmentation.

\end{abstract}
\section{Introduction}
Automatic organ segmentation is critical in analyzing human anatomy. In addition to a qualitative assessment by the radiologist, quantitative assessment using automatic segmentation assists in the accurate diagnosis of diseases. Multi-organ segmentation can assess organ interactions and provide more detailed information than single-organ segmentation. Although multiorgan segmentation plays a vital role in computer-aided diagnosis, it suffers from computational-memory limitations. Moreover, because CT and MRI images acquired in daily clinical practice comprise high-resolution 3D images, this limitation makes it even more difficult to input whole images into deep learning models.
Current methods usually resize images to smaller sizes at the expense of organ details or crop a portion of the image (patch) and input it into the model without using the positional information of the surrounding organs. Some studies have shown that a large patch size allows a model to make more accurate predictions \cite{Hamwood2018-zu,Imai2018-ls}.

The batch size is also a critical parameter that influences training effectiveness. Larger batch sizes are associated with more accurate gradient estimates \cite{Isensee2021-dl}; however, batch size is also affected by memory limitations. In this study, the model could only be trained with a smaller batch size for the 3D organ segmentation compared to tasks, such as natural image recognition. 

Therefore, we hypothesize that increasing the patch and batch sizes would improve the training accuracy. To test this hypothesis, we used an nnU-Net \cite{Isensee2021-dl} model, which is the gold standard for medical image segmentation. The dataset used was the Abdominal Multi Organ Segmentation 2022 (AMOS2022) challenge \cite{ji2022amos}, which consists of 500 CT and 100 MRI scans collected from multiple sites, a wide range of imaging conditions, and patient backgrounds with 15 voxel-level annotations. We trained the model with distributed learning using multiple GPUs and searched for the batch and patch sizes with the best segmentation accuracy.

\section{Method}

\subsection{Approach}

In this study, the preprocessing, network architecture, training strategy, and postprocessing were based on the default nnU-Net configuration.
As referred to in the original paper, a larger batch size enables more accurate gradient estimates, and a larger patch size increases contextual information among organs \cite{Isensee2021-dl}. Loss function is another factor that significantly affects the accuracy of the training process. We compared changes in accuracy by varying these parameters.

\subsubsection{Training with Large Patch Size}\mbox{}\\
nnU-Net initializes the patch size to the median image shape and iteratively reduces it while adapting the network topology accordingly, until the network can be trained with a batch size of at least 2. For example, on an NVIDIA RTX3090, the patch size was automatically set to (depth,height,width) = (80,160,160) for both Task1 and Task2 by the default nnU-Net configuration. We set the patch size to [64,160,160], [64,288,288], [64, 354,354], and [64,384,384]. In all cases, the batch size was set to 2.

\subsubsection{Training with Large Batch Size}\mbox{}\\
Multi-organ segmentation using a 3D deep learning model is often performed with a batch size of 1-4\cite{Feng2021-fz,Tang2021-pl,Thaler2021-cs,Kakeya2018-hk,Muller2021-mw}, which is smaller than the batch size used for learning typical 2D images (usually 16 or more)\cite{Hollon2020-ei,Irvin2019-us}. The default nnU-Net configuration requires a batch size of at least 2. In our experiments, we varied the batch size from 2 to 16.

\subsubsection{Loss Function}\mbox{}\\
Compound loss allows for robust predictions in medical image segmentation\cite{Ma2021-kc}. In this study, we combined dice, cross-entropy (CE), focal\cite{Lin2017-lc}, top-k\cite{Lapin2016-vj}, and perimeter loss\cite{Jurdi2021-mo}, which are commonly used for segmentation, to search for the best compound loss in this challenge.

\subsection{Dataset}
AMOS 2022 contains two tasks: 1) a CT-only task and 2) a cross-modal task between CT and MRI. In Task1, we used 200 CT cases for training and 100 cases for validation. In Task2, we used 200 CT and 40 MRI cases for training and 100 CT and 20 MRI cases for validation. In our experiments, accuracy was evaluated on the Task2 dataset. Based on these results, we set the batch size, patch size, and loss function for the final Task1 and Task2 submissions.

\subsection{preprocessing}
Resampling, data transformation, and normalization were performed according to the nnU-Net configuration. All cases were resampled to a common voxel spacing of 2.0 × 0.68825001 × 0.68825001 in Task 1, and 2.0 × 0.78014851 × 0.78014851 in Task 2.

\subsection{Postprocessing and Ensemble}
For the final submission, we used an ensemble of two models trained using different patch sizes, batch sizes, and loss functions. Each model was an independent ensemble of its five trained cross-validation folds. Postprocessing was applied to each model’s output before the ensemble.

\subsection{Implementation Details}
Multi-GPU distributed learning was used to train the model. The network was trained using an NVIDIA RTX3090 and SQUID (Supercomputer for Quest to Unsolved Interdisciplinary Datascience, Osaka University) with an NVIDIA A100 40 GB × 8 GPU. The main component of our framework was 3D U-Net. The initial learning rate was set to 0.01. The optimizer is stochastic gradient descent with Nesterov momentum ($\mu=0.99$). Segmentation performances were calculated by one-fold when each parameter was changed, and the final result was an ensemble of models with 5-fold cross-validation.

\section{Result}
Based on the request from the organizers, two metrics were used for evaluation: dice similarity coefficient (DSC) and normalized surface dice (NSD). Table 1 shows the change in accuracy with patch-size variations. The DSC and NSD scores tended to increase as the patch size increased, but there was almost no increase in accuracy for patch sizes larger than 300. For patch sizes of (64,384,384), both the DSC and NSD scores worsened significantly.

\begin{table}[]
\caption{Performance of different patch size settings.}
\centering
\begin{tabular*}{10cm}{@{\extracolsep{\fill}}ccc}
\toprule
Patch Size     & DSC Score & NSD Score \\ \midrule
(64, 160, 160) & 0.8926    & 0.8138    \\
(64, 288, 288) & 0.8972    & \textbf{0.8256}    \\
(64, 354, 354) & \textbf{0.8973}    & 0.8255    \\
(64, 384, 384) & 0.8715    & 0.7731    \\ \bottomrule
\end{tabular*}%
\end{table}

\begin{figure}
\includegraphics[width=\textwidth]{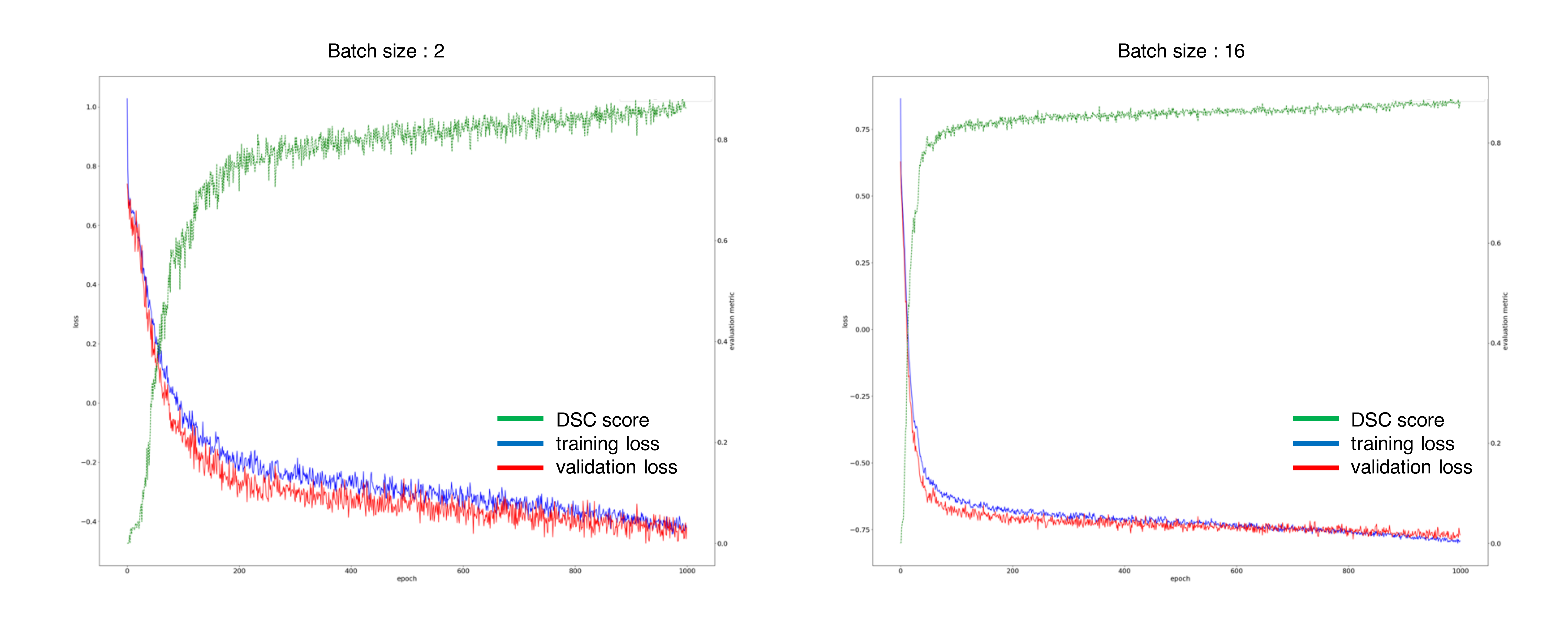}
\caption{Transition for losses and DSC score in batch size=2 and 16.} \label{fig1}
\end{figure}

Next, Table 2 shows the change in accuracy when batch size was changed. Accuracy increased as the batch size increased, but it decreased slightly when the batch size was increased to 64. The transition of training loss and DSC score by batch size is shown in Fig.~\ref{fig1}, indicating that a large batch size enabled a more stable learning progress.

\begin{table}[]
\caption{Performance of different batch size settings.}
\centering
\begin{tabular*}{10cm}{@{\extracolsep{\fill}}ccc}
\toprule
Batch Size & DSC Score       & NSD Score       \\  \midrule
2          & 0.8926          & 0.8138          \\
4          & 0.8950          & 0.8202          \\
16         & \textbf{0.9040} & \textbf{0.8360} \\
64         & 0.9027          & 0.8344          \\ \bottomrule
\end{tabular*}
\end{table}

Additionally, the loss function was compared to determine the loss with the highest score. The batch size was set to 2, and the patch size was (64, 288, 288). As shown in Table 3, dice and CE loss had the highest accuracy in NSD, while dice and focal loss had the best DSC score.

\begin{table}[]
\caption{Performance of different composite loss functions.}
\centering
\begin{tabular*}{10cm}{@{\extracolsep{\fill}}lcc}
\toprule
Loss Function              & DSC Score       & NSD Score       \\ \midrule
Dice \& CE Loss*\tnote{1} & 0.8972          & \textbf{0.8256} \\
Dice \& Focal Loss         & \textbf{0.8981} & 0.8250          \\
Dice \& TopK Loss          & 0.8952          & 0.8221          \\
Dice \& Perim Loss         & 0.8949          & 0.8204          \\ \bottomrule
\end{tabular*}
\begin{tablenotes}
\item[1]\hspace*{2em}* nnU-Net default loss function
\end{tablenotes}
\end{table}

\begin{figure}
\includegraphics[width=\textwidth]{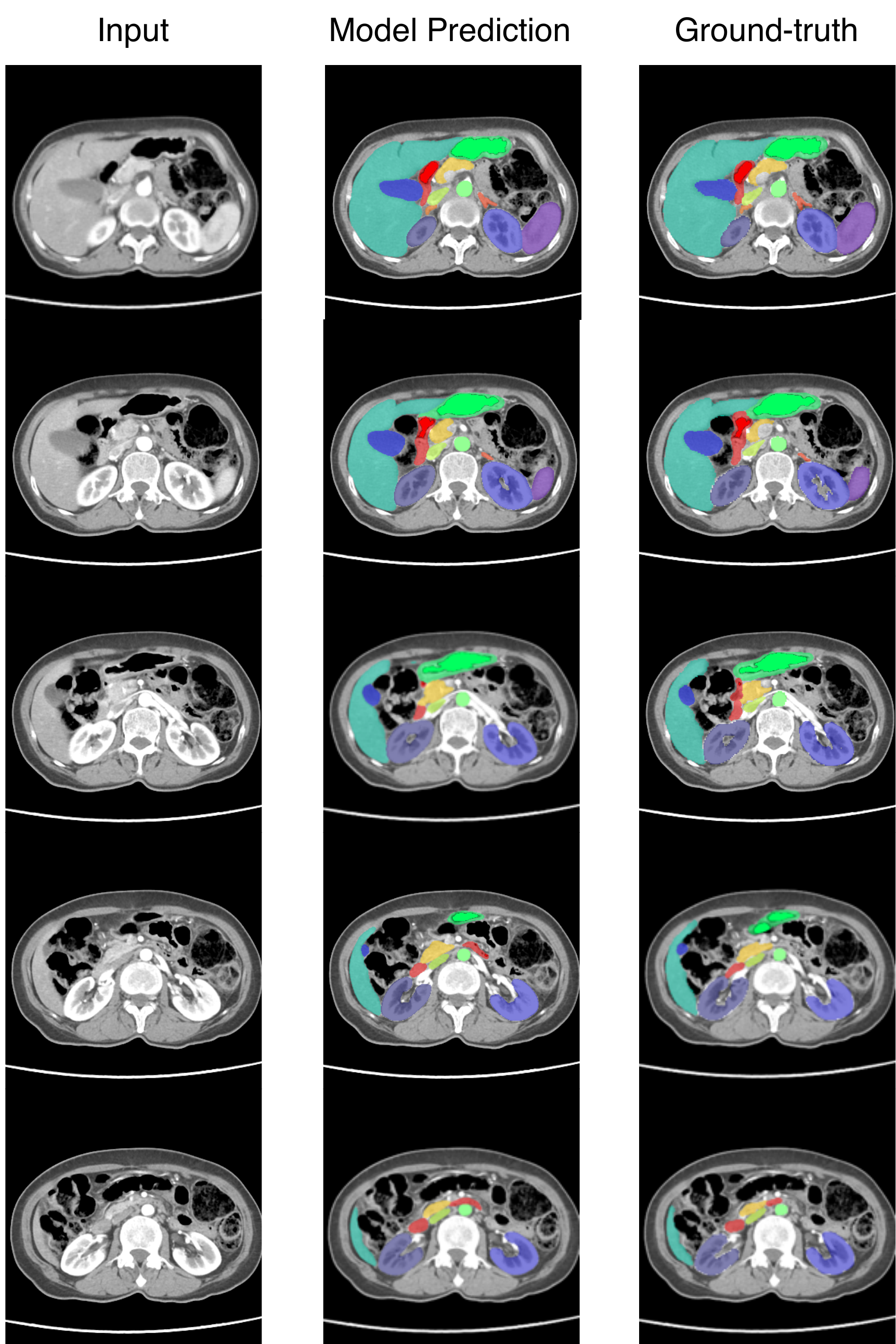}
\caption{Validation examples of our final model in Task 1. Left: axial view of input contrast-enhanced CT images. They are shown in order by 10mm slices. Middle: the predictions of our final ensemble model. Right: the ground-truth labels.} \label{fig2}
\end{figure}

\begin{figure}
\includegraphics[width=\textwidth]{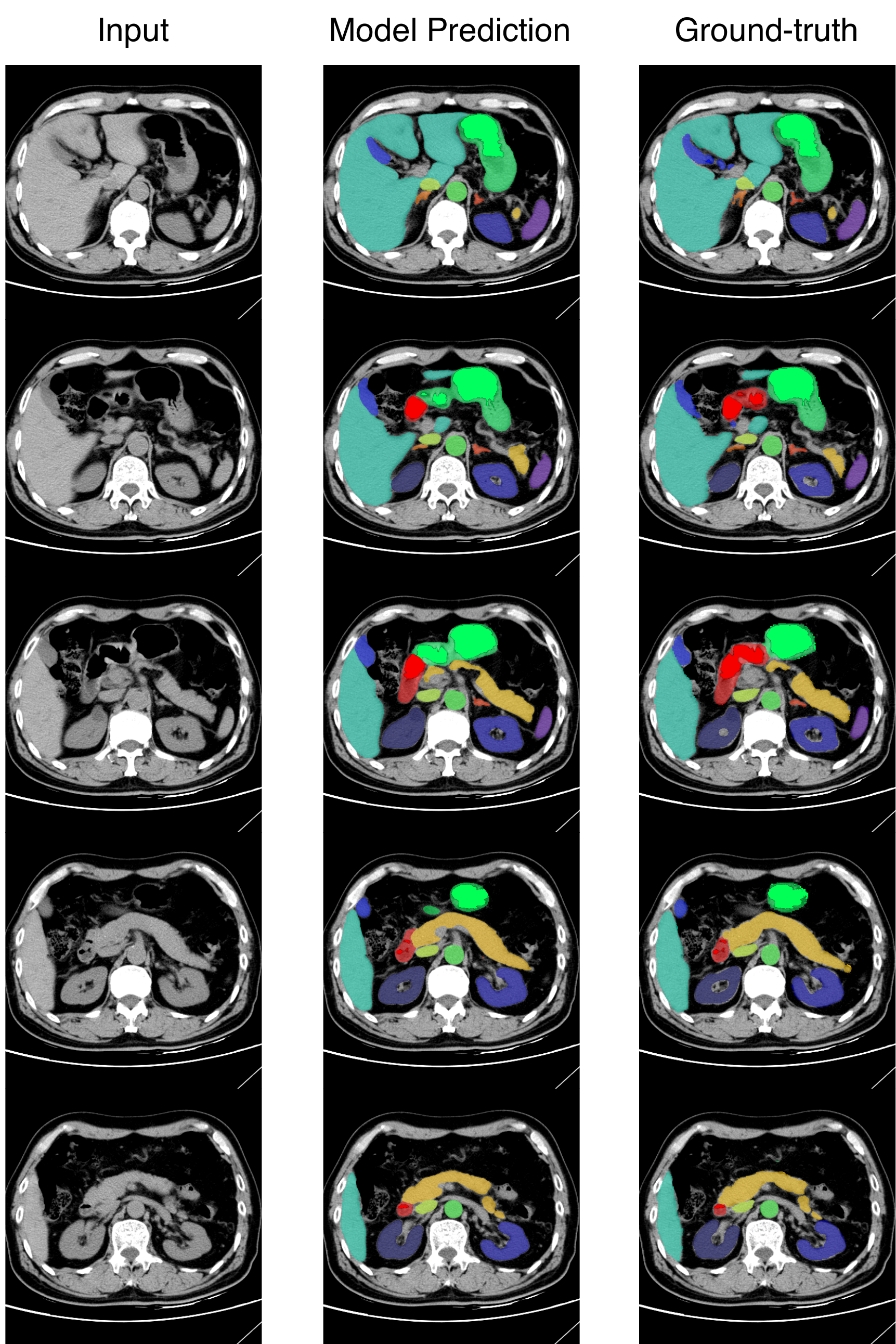}
\caption{Validation examples of our final model in Task 1. Left: axial view of input non-contrast-enhanced CT images. They are shown in order by 10mm slices. Middle: the predictions of our final ensemble model. Right: the ground-truth labels.} \label{fig3}
\end{figure}

Based on these results, an ensemble of model A (batch size=16, patch size (64,160,160), loss=dice and CE) and model B (batch size=8, patch size (64,288,288), loss=dice and focal) was selected as the final model. The results of these ensembles are shown in Table 4. Task 1 exhibited an accuracy of 0.9139 for DSC and 0.8459 for NSD, whereas Task 2 had an accuracy of 0.9089 for DSC and 0.8453 for NSD. Examples of the prediction results are shown in fig~\ref{fig2} and fig~\ref{fig3} for contrast and non-contrast, respectively.

\begin{table}[]
\caption{Performance of different training settings and ensembles. All the score is the ensemble of 5-fold cross validation models.}
\centering
\begin{tabular}{llcc}
\toprule
                        & Model (5-fold cross validation)                      & DSC Score       & NSD Score       \\ \midrule
Task 1 & model A (64,160,160), bs=16, Dice\&Crossentropy loss & 0.9113          & 0.8417          \\
                        & model B (64,288,288), bs=8, Dice\&Focal loss         & 0.9127          & 0.8450          \\
                        & ensemble(model A \& B)                               & \textbf{0.9139} & \textbf{0.8459} \\ \midrule
Task 2 & model A(64,160,160), bs=16, Dice\&Crossentropy       & 0.9064          & 0.8412          \\
                        & model B (64,288,288), bs=8, Dice\&Focal loss         & 0.9080          & 0.8443          \\
                        & ensemble(model A \& B)                               & \textbf{0.9089} & \textbf{0.8453} \\ \bottomrule
\end{tabular}
\end{table}

\section{Discussion}

In this study, we examined changes in accuracy with patch and batch sizes larger than the default nnU-Net settings. Our results showed that larger patch and batch sizes tended to result in higher accuracy. In addition, ensemble models trained with different parameters improved segmentation performance. However, this study had some limitations: other parameters, such as learning rate, optimizer, and network architecture, were not changed. In particular, batch size exhibited a significant relationship with the learning rate \cite{l.2018dont}. We will experiment with other parameters including the learning rate to create a model with higher accuracy. Moreover, the results of this study showed no significant improvement in accuracy when the batch size exceeded 16 or the patch size exceeded 288, indicating that increasing the patch and batch size did not always improve accuracy. These results may serve as indicators for setting parameters for medical image segmentation. In the future, we hope to develop a computationally resource-efficient learning method that can capture a wide range of areas in an image and achieve stable learning progress.

%
%
%
\bibliographystyle{splncs04}
\bibliography{reference}

\end{document}